\documentclass{ws-rv9x6}
\usepackage{subfigure}   
\makeindex

\begin{document}

\chapter[]{Final state multiplicity and particle correlation in small systems} 

\author[V.Mariani]{V.Mariani\footnote{On behalf of the CMS Collaboration}}

\address{Universit\'a degli Studi di Perugia, IT\\
vmariani@cern.ch}

\begin{abstract}
Final state variables and particle correlation will be discussed under a Multiple Parton Interaction (MPI) interpretation.
The state of the art about the latest results on such variables will be provided.  
Furthermore the role played by event multiplicity in the deep understanding of particle correlation, in particular concerning the new results on the Long-Range Near-Side two particle correlations by the CMS Collaboration, will be discussed.
\end{abstract}

\body

\section{Introduction}\label{intro}

Historically the final state multiplicity plays a key role in Multiple Parton Correlation studies.
Most of the Monte Carlo (MC) generators in fact are not able to describe simultaneously both energy evolution as a function of pseudorapidity, $\eta$, and transverse momentum, $p_T$.
Only when the hypothesis of Multiple Parton Interactions (MPIs) is introduced in the models a better description of data is performed.  
In order to have a complete picture of the current MC modelling status, in section \ref{FSM} the most recent analysis about pseudorapidity and transverse momentum distributions for charged particles, hadronic event shape and forward energy measurement will be shown and discussed with a comparison between different collaboration results and center of mass energy values. 
Since multiplicity plays a key role in the Long-Range Near-Side (LRNS) correlation, in section \ref{LRNS} the results from the CMS Collaboration show that a ridge-like structure, seen previously in proton-nuclei (p-A) and nuclei-nuclei (A-A) collisions, appears in proton-proton (p-p) collision events only when we look at high multiplicity events.

\section{Final State Multiplicity}\label{FSM}

A description of the most characteristic final state multiplicity variables will follow.

\subsection{Pseudorapidity and Transverse Momentum distributions of charged particles}

The measurements of the charged particle yields and kinematic distributions are essential in exploiting the energy regimes of particle collisions at the LHC.
The charged particles pseudorapidity distribution and the charged particles $p_T$ distribution can be defined as in \cite{leading_transverse_momentum_distributions_of_charged_particles_CMS}.

\begin{figure}
\centering
\includegraphics[height=0.38\textwidth]{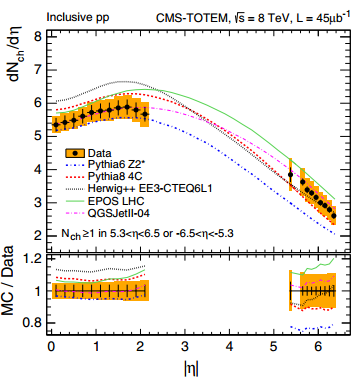}
\includegraphics[height=0.38\textwidth]{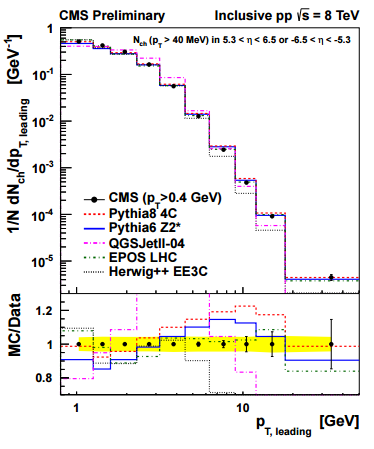}
\caption{Figure on the left shows the pseudorapidity distribution $dN_{ch}/d\eta$ obtained by the CMS and TOTEM collaborations at $\sqrt{s} = 8$ TeV \cite{pseudorapidity_distributions_of_charged_particles_CMSTOTEM}. Figure on the right shows the momentum distribution $dN_{ch}/dp_T$ obtained by the CMS Collaboration at $\sqrt{s} = 8$ TeV \cite{leading_transverse_momentum_distributions_of_charged_particles_CMS}. Both the distributions are compared to different Monte Carlo models}
\label{CMSpiuTOTEM}
\end{figure}

%
%
%
In Fig.\ref{CMSpiuTOTEM} (left) one can see the pseudorapidity distribution ($dN_{ch}/d\eta$) obtained at $\sqrt{s} = 8$ TeV by the CMS and TOTEM collaborations \cite{pseudorapidity_distributions_of_charged_particles_CMSTOTEM} analysing the pseudorapidity ranges $|\eta| < 2.2$ and $5.3 < |\eta| < 6.4$.
In Fig.\ref{CMSpiuTOTEM} (right) the momentum distribution ($dN_{ch}/dp_T$) from the CMS Collaboration \cite{leading_transverse_momentum_distributions_of_charged_particles_CMS} for charged particles in the pseudorapidity ranges $|\eta| < 2.4$ is shown. 

Monte Carlo models tuned to Underlying Event observables provide the best description of data for both the observables in Fig.1.

\begin{figure}
\centering
\includegraphics[height=0.38\textwidth]{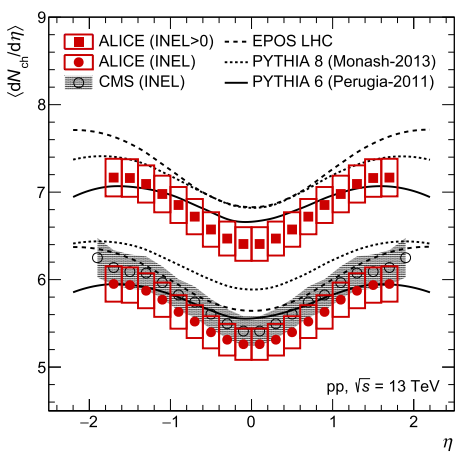}
\caption{The average of the pseudorapidity function is shown by the ALICE Collaboration comparing data at $\sqrt{s} = 13$ TeV to the Monte Carlo calculations and the results from the CMS Collaboration \cite{pseudorapidity_distributions_of_charged_particles_ALICE}.}
\label{ALICE}
\end{figure}

An interesting analysis on the pseudorapidity distribution of charged particles was done by the ALICE Collaboration at 13 TeV \cite{pseudorapidity_distributions_of_charged_particles_ALICE}.
The pseudorapidity distribution is measured at central rapidity in $|\eta| < 1.8$. The measurements reported have been obtained for inelastic events (INEL) and events having at least one charged particle produced with $p_T$ $>$ 0 in the pseudorapidity interval $|\eta| < 1$ (INEL $>$ 0). They compare their results to the CMS ones \cite{pseudorapidity_distributions_of_charged_particles_CMS} and to different Monte Carlo models as shown in Fig.\ref{ALICE}.

The results from the two collaborations agree within the uncertainties; it is shown that PYTHIA 6 (with Perugia tune \cite{PerugiaTune} based on MPIs) calculations is in better agreement with data than PYTHIA 8 for both selections.

%

\begin{figure}
\centering
\includegraphics[height=0.38\textwidth]{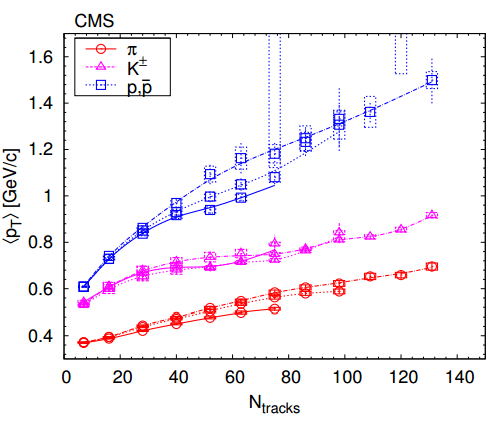}
\caption{The average of the $p_T$ values is shown as a function of the event multiplicity for three different particle species: pions (red), kaons (magenta) and protons (blue) \cite{inclusive_production_of_charged}.}
\label{pt_mult}
\end{figure}

The CMS Collaboration also studied the dependence of the average of the $p_T$ from the event charged-particle multiplicity for different identified particles at $\sqrt{s}$ = 0.9, 2.76, and 7 TeV \cite{inclusive_production_of_charged}: pions, kaons and protons.
The distribution in Fig.\ref{pt_mult} shows that the correlation between $<p_T>$ and the event multiplicity is not negligible.

Assuming MPI can help the interpretation: a higher multiplicity is correlated with more multiple parton interactions, hence, an higher $p_T$ average value is expected.

\subsection{Hadronic Event Shape}

A hadronic event shape can be described by three common variables.
\begin{itemize}
\item{\bf{Transverse thrust:}} $\tau_{\bot} = 1 - max_{\hat{\eta}_T}{{\sum_i|\vec{p}_{T,i}\cdot\hat{\eta}_T}|\over{\sum_i\vec{p}_{T,i}}}$, $\tau_{\bot}$ = 0 for perfectly balanced two-jet events and $\tau_{\bot} = (1-2/\pi)$ in isotropic multijet events   
\item{\bf{Sphericity} and \textbf{Transverse Sphericity:}} $S = {3\over2}(\lambda_2 + \lambda_3)$ and $S_{\bot} = {2\lambda_2 \over {\lambda_1 + \lambda_2}}$ where $\lambda_1$, $\lambda_2$ and $\lambda_3$ are the normalized eigenvalues ($\lambda_1 < \lambda_2 < \lambda_3$) of the momentum tensor.
\end{itemize}

\begin{figure}
\centering
\includegraphics[height=0.31\textwidth]{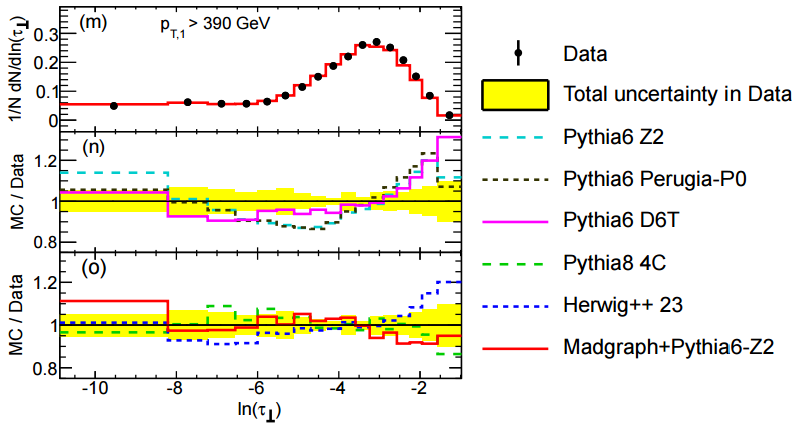}
\includegraphics[height=0.31\textwidth]{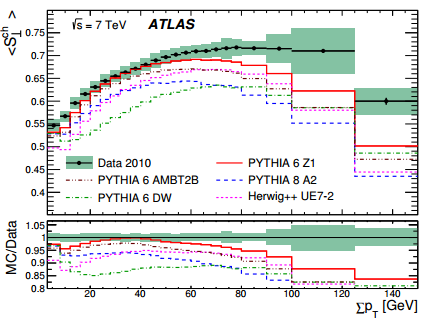}
\caption{Figure on the left shows the transverse thrust distribution with CMS data collected at $\sqrt{s}$ = 7 TeV \cite{hadronic_event_shape_CMS}. Figure on the right shows the sphericity distribution as a function of $p_T$ obtained by the ATLAS Collaboration at $\sqrt{s} = 7$ TeV \cite{hadronic_event_shape_ATLAS}. Both the distributions are compared to different Monte Carlo models}
\label{event_shape}
\end{figure}

In Fig.\ref{event_shape} the transverse thrust distribution and the sphericity are shown by the CMS \cite{hadronic_event_shape_CMS} and ATLAS \cite{hadronic_event_shape_ATLAS} collaborations respectively.

Events with a large number of MPIs are expected to be isotropic, i.e. should appear with a spherical shape, especially for high multiplicity regime. 
MPIs are correlated to the multiplicity, in this sense final states with higher $<p_T>$ are produced. 
In fact from these distributions it is shown that transverse thrust describes an higher isotropic contribution than expected in jet events and the sphericity is higher in high $p_T$ (and high multiplicity) events than expected. The disagreement between data and Monte Carlo increases at large $\sum p_T$ values. 

\subsection{Forward Energy Measurement}

\begin{figure}
\centering
\includegraphics[height=0.38\textwidth]{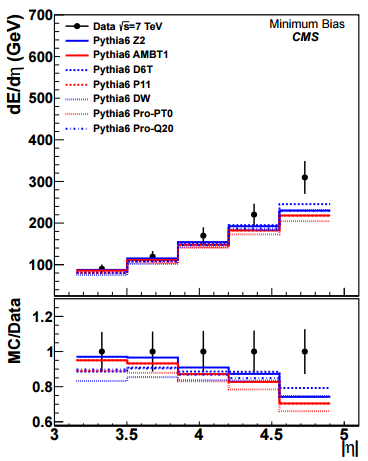}
\includegraphics[height=0.38\textwidth]{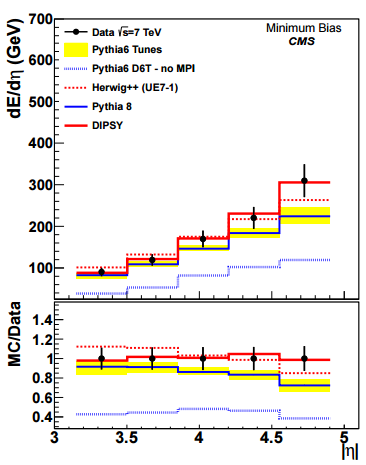}\\
\includegraphics[height=0.38\textwidth]{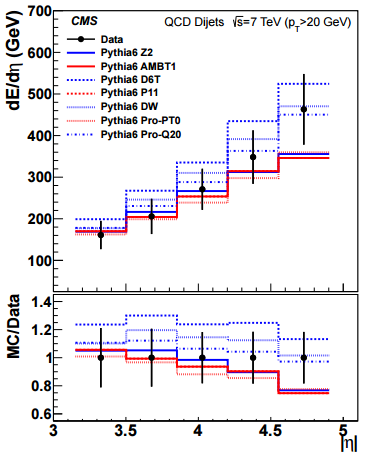}
\includegraphics[height=0.38\textwidth]{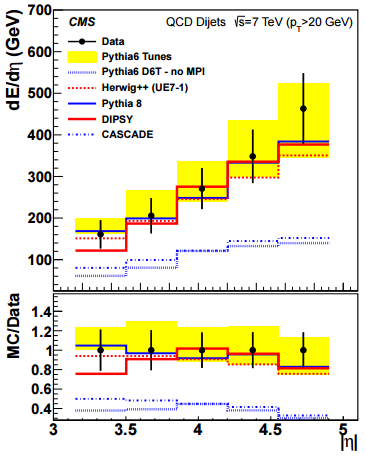}
\caption{The energy spectrum in the forward region, $3.15 < |\eta| < 4.9$ is shown for data collected by the CMS experiment at $\sqrt{s} =$ 7 TeV with two different triggers: Minimum Bias (top) and Di-jets (bottom) \cite{forward_energy_spectrum_7tev}. }
\label{energy_spectrum_7tev}
\end{figure}

In order to factorize the UE observables from the Beam-Beam Remnant (BBR) contribution it is useful to study the MPI in the forward region.
The last important measurement concerning the final state multiplicity variables to point out is the forward energy spectrum done by the CMS Collaboration at $\sqrt{s} =$ 7 \cite{forward_energy_spectrum_7tev} and 13 TeV \cite{forward_energy_spectrum_13tev}, \cite{forward_energy_spectrum_13tev_2015}; the two analysis can be considered complementary hence it is interesting to mention both of them.
The former measures the energy flow with the hadronic forward calorimeters in $3.15 < |\eta| < 4.9$, for events selected with different triggers.
In Fig.\ref{energy_spectrum_7tev} $dE/d\eta$ distributions are shown for 7 TeV data selected with Minimum Bias trigger (top) and di-jet trigger (bottom) with a comparison to several Monte Carlo models.

The energy flow increases with the pseudorapidity, as expected, and the comparison between these two samples is needed to outline the role played by the event multiplicity.
In the events triggered by a Jet trigger, where the track multiplicity is higher, the energy flow is significantly higher than the Minimum Bias events and the discrepancy between data and MC models is larger.
It is important to underline that MPI are unavoidable to best reproduce data in the forward region.

\begin{figure}
\centering
\includegraphics[height=0.38\textwidth]{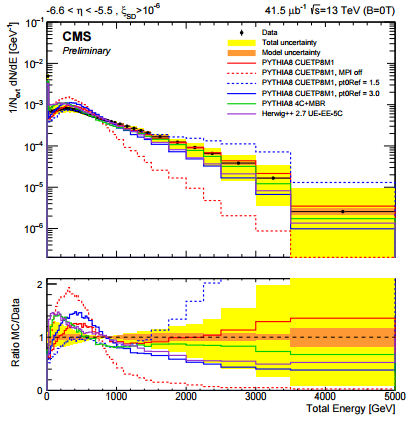}
\caption{The energy spectrum in the very forward region, $-6.6 < \eta < -5.2$ is shown for data collected by the CMS experiment at $\sqrt{s} =$ 13 TeV \cite{forward_energy_spectrum_13tev}.}
\label{energy_spectrum_13tev}
\end{figure}

The analysis done at 13 TeV \cite{forward_energy_spectrum_13tev} measures the event-by-event energy deposition into the very forward acceptance, $-6.6 < \eta < -5.2$, using data from the electromagnetic and hadronic sections of the CASTOR calorimeter (a wider pseudorapidity range is studied in \cite{forward_energy_spectrum_13tev_2015}).
Results are shown in Fig.\ref{energy_spectrum_13tev} where data are compared to several MC models and tunes. None of these models consistently describes the shape in both low and high multiplicity. PYTHIA8 CUETP8M1, based on Underlying Event observables and MPI theory, seems to provide the best description.
Even in this case the prediction of models that not include MPI is ruled out by the data.
Furthermore it is shown that data are very sensitive to the MPI cut-off. 

\section{Particle Correlation}\label{LRNS}

The two-particle angular correlations for charged particles are studied in both short range, $|\Delta\eta < 2|$, and long range, $2 < |\Delta\eta| < 4.8$.
Defining:
\begin{itemize}
\item{Signal function} $S_N (\Delta\eta,\Delta\Phi) = {{1}\over{N(N-1)}} {{d^2N^{sign}}\over{d\Delta\eta\Delta\Phi}}$ as two-particle pair correlated in the same events 
\item{Background function} $B_N(\Delta\eta,\Delta\Phi)= {{1}\over{N^2}}{{d^2N^{mixed}}\over{d\Delta\eta\Delta\Phi}}$ as the distribution of uncorrelated particle pairs from two randomly selected events
\end{itemize}

the correlation function is defined as 

\begin{equation}
R(\Delta\eta\Delta\Phi)= \left \langle(\left \langle N \right \rangle -1)\left( {{S_N(\Delta\eta,\Delta\Phi)}\over{B_N(\Delta\eta,\Delta\Phi)}} -1 \right)\right\rangle_{bins}
\end{equation}

\subsection{Long-Range Near-Side Two particle angular correlation at 13 TeV}

\begin{figure}
\centering
\includegraphics[height=0.38\textwidth]{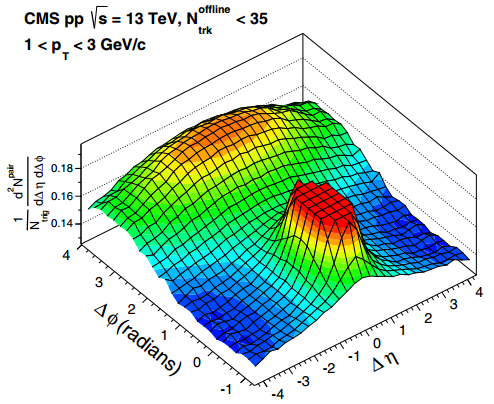}
\includegraphics[height=0.38\textwidth]{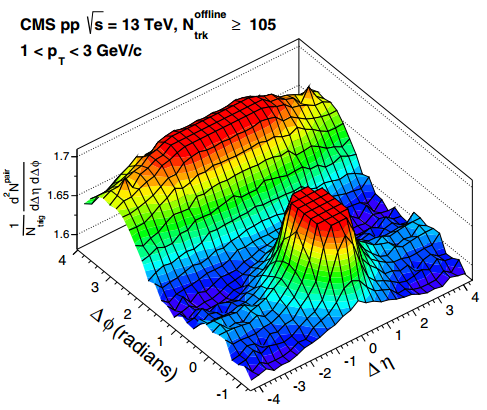}
\caption{The 2D $(\Delta\eta, \Delta\Phi)$ two-particle correlation functions in pp collisions at $\sqrt{s} = $ 13 TeV for pairs of charged particles both in the range 1 $<$ $p_T$ $<$ 3 GeV/c is shown \cite{LRNS}. Results are shown for low multiplicity events ($N^{offline}_{trk} < 35$) on the left and for high-multiplicity events ($N^{offline}_{trk} \ge 105$) on the right. The sharp peaks from jet correlations around $(\Delta\eta, \Delta\Phi) = 0 $ are truncated to better illustrate the long-range correlations.}
\label{LRNS_peak}
\end{figure}

The latest results on the Long-Range Near-Side two-particle correlation by the CMS Collaboration are reported in the paper \cite{LRNS}. 
It is shown that analysing charged particles in the $p_T$ range $1 < p_T < 3$ GeV two very different behaviours appear depending on the multiplicity.
For a low multiplicity sample, i.e. $N_{trk}^{offline} < 35$, the dominant features are the peak near $(\Delta\eta,\Delta\Phi) = (0,0)$ for particle pairs within the same jet and the elongated structure at $\Delta\Phi \approx \pi$ for recoiling activity as show in Fig.\ref{LRNS_peak} (left).
Looking at high multiplicity events, i.e. $N_{trk}^{offline} > 105$, in addition to these jet-like correlation structures, a "ridge"-like structure appears at $\Delta\Phi \approx 0$ extending over a range of at least 4 units in $|\Delta\eta|$, see Fig.\ref{LRNS_peak} (right).

This evidence confirms what was already observed at $\sqrt{s} = 7$ TeV \cite{LRNS7tev} for the first time in proton-proton collisions, even if no such long-range correlations are predicted by PYTHIA. 

\begin{figure}
\centering
\includegraphics[height=0.38\textwidth]{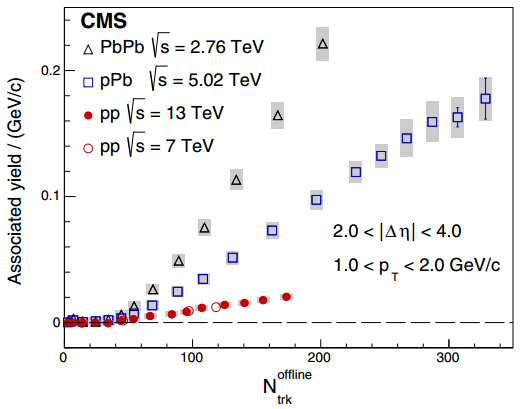}
\caption{Associated yield of long-range near-side two-particle correlations for 1 $<$ $p_T$ $<$ 2 GeV/c in pp collisions at $\sqrt{s} =$ 13 and 7 TeV, pPb collisions at $\sqrt{s_{NN}} =$ 5.02 TeV, and PbPb collisions at $\sqrt{s_{NN}} =$ 2.76 TeV. 
Associated yield for the near side of the correlation function is averaged over 2 $< |\Delta\eta| <$ 4 \cite{LRNS}.}
\label{LRNS_ss}
\end{figure}

A yield can be associated to the LRNS structure as discussed in \cite{LRNS}; this corresponding yield has been measured for proton-proton (p-p), proton-lead (p-Pb) and lead-lead (Pb-Pb) collisions in CMS, in Fig.\ref{LRNS_ss} the associated yield is reported as a function of the track multiplicity.

The ridge-like correlations become significant at a multiplicity value of about 40 in all the three systems and exhibit a nearly linear increase for higher value.
For a given multiplicity value the associated yield in pp collision is roughly 10 $\%$ and 25 $\%$ of those observed in pPb and PbPb collisions respectively.
There is a strong collision system size dependence of the correlations while the dependence of the center of mass energy seems to be negligible.

There are many interpretations available to explain the ridge-effect, in particular:
\begin{itemize}
\item{\bf{Hydrodynamic models}} two- and multi-particle angular correlations in pp collisions have been studied as a function of charged-particle multiplicity by the CMS Collaboration \cite{collectivity}. The second-order (v2) and third-order (v3) azimuthal anisotropy harmonics of unidentified charged particles, as well as v2 of $K^0_s$ and $\Lambda/\bar{\Lambda}$ particles, are extracted from long-range two-particle correlations as functions of particle multiplicity and transverse momentum. 
For high-multiplicity pp events, a mass ordering is observed for
the v2 values of charged hadrons (mostly pions), $K^0_s$, and $\Lambda/\bar{\Lambda}$, with lighter particle species exhibiting a stronger azimuthal anisotropy signal below $p_T$ $\approx$ 2 GeV/c. For
13 TeV data, the v2 signals are also extracted from four- and six-particle correlations for the first time in pp collisions, with comparable magnitude to those from two-particle correlations. These observations are similar to those seen in pPb and PbPb collisions, and support the interpretation of a collective origin for the observed long range correlations in high-multiplicity pp collisions.
\item {\bf{MPIs}} since Pythia models can't reproduce the ridge-like structure, in \cite{LRNS_mpi} a modification of Pythia6 has been proposed, explaining the ridge effect with Multiple Parton Interactions.
The model introduces a correlation between the azimuth of the event planes of individual MPI and the event plane of the hardest interaction. 
This correlation can be naturally explained in a physical picture based on the impact parameter between the protons. 
The two main implications of this modification are the appearance of the near-side ridge in high-multiplicity moderate-$p_T$ events and a shift in the activity in the transverse region. 
\end{itemize}

\section{Conclusion}

The latest results on final state variables and particle correlation studies have been presented under a Multiple Parton Interaction (MPI) interpretation.
MPI models are unavoidable to have a complete description of the LHC final states and to explain the high multiplicity events in the correlation effects.
It is also clear that multiplicity plays a key role in the final state, but this is still not completely understood, as shown by the large deviations between data and MC in several results in high multiplicity events, and in the particle correlation as well.
There are many hypotheses to explain the origin of the LRNS correlation (hydronamic model, MPI, Colour Glass Condensate etc..) but a deep investigation is still needed.

\end{document}